\documentclass[aps,pra,showpacs,floatfix,superscriptaddress,twocolumn]{revtex4}
\usepackage{graphicx}
\usepackage{subfig}
\usepackage{amssymb,amsmath,amsfonts,mathrsfs,eufrak,fancyhdr}
\usepackage{hyperref}
\usepackage{color}
\usepackage{makeidx,endnotes}
\usepackage{showidx}
\usepackage{theorem,eepic}
\usepackage{epsf,epsfig}

\def\<{\langle}
\def\>{\rangle}

\def\set#1{{\sf #1}}

\def\map#1{{\mathcal{#1}}}
\def\hil#1{{\mathscr{#1}}}

\def\Tr{\operatorname{Tr}}

\newtheorem{lemma}{Lemma}[section]

\newtheorem{corollary}[lemma]{Corollary}

\newtheorem{theorem}[lemma]{Theorem}

\def\Proof{\medskip\par\noindent{\bf Proof. }}

\begin{document}
\title{Asymptotic properties of quantum Markov chains}
\author{Jaroslav Novotn\'y}
\affiliation{Department of Physics, FNSPE, Czech Technical University in Prague, B\v rehov\'a 7, 115 19 Praha 1 - Star\'e M\v{e}sto, Czech Republic}
\affiliation{Institut f\"ur Angewandte Physik, Technische Universit\"at Darmstadt, D-64289 Darmstadt, Germany}
\author{Gernot Alber}
\affiliation{Institut f\"ur Angewandte Physik, Technische Universit\"at Darmstadt, D-64289 Darmstadt, Germany}
\author{Igor Jex}
\affiliation{Department of Physics, FNSPE, Czech Technical University in Prague, B\v rehov\'a 7, 115 19 Praha 1 - Star\'e M\v{e}sto, Czech Republic}

\date{\today}

\begin{abstract}
The asymptotic dynamics of quantum Markov chains generated by
the most general physically relevant quantum operations is investigated. It is shown that it
is confined to an attractor space on which the resulting quantum Markov chain is diagonalizable.
A construction procedure of a basis of this attractor space and its associated dual basis is presented. It applies whenever
a strictly positive quantum state exists which is contracted or left invariant by the generating quantum operation. Moreover, algebraic relations between the attractor space and Kraus operators involved in the definition of a quantum Markov chain are derived.
This construction is not only expected to offer significant computational
advantages in cases in which the dimension of the Hilbert space is large
and the dimension of the attractor space is small but
it also sheds new light onto the relation between the asymptotic dynamics of quantum Markov chains and fixed points of their
generating quantum operations.
\end{abstract}


\maketitle

\section{Introduction}
Quantum operations, i.e. completely positive and trace non-increasing linear transformations acting
within a  Hilbert space, play a central role in quantum theory. They
describe the most general dynamics of an open quantum system which can be induced
by unitary transformations and von Neumann measurements involving an additional
initially uncorrelated ancillary quantum system
 \cite{quantumoperations}.
In particular, in quantum information theory quantum operations
are an indispensable theoretical tool for exploring the ultimate limits of quantum theory \cite{possible+impossible}.

Recent quantum technological advances \cite{expadvances}
have stimulated significant interest in the dynamics of large
open quantum systems which are formed by many either indistinguishable
or distinguishable elementary quantum systems, such as Bose-Einstein
condensates \cite{BEC} or qubit-based quantum networks \cite{Networks}.
These technological developments raise interesting theoretical questions concerning the
dynamics of large quantum systems under the action of iterated quantum
operations, so called quantum Markov chains \cite{quantumMarkovdefinition}.
This is due to the fact that many discrete models in the area of
statistical physics by which fundamental physical
phenomena, such as the approach to thermal equilibrium or the asymptotic dynamics and decoherence
of macroscopic quantum systems, are explored involve quantum Markov chains.

For a general classification of quantum  Markov chains
an investigation of their asymptotic dynamics
resulting from large numbers of iterations of their generating quantum operations constitutes a natural starting point.
In this context the natural questions arise
which asymptotic dynamics is possible for a quantum Markov chain and how it is related to
spectral properties of its generating quantum operation. It is a main intention of this paper to address these questions
for arbitrary quantum Markov chains.

Recently,
some results addressing these questions
have already been found
for special classes of quantum Markov chains.
In particular, the asymptotic dynamics of Markov chains resulting from iterated random unitary quantum operations
has been
investigated in detail. These operations are contracting and hence
two major results have been established \cite{NovotnyJPA,NovotnyCEJP}. Firstly, it has been demonstrated that the asymptotic dynamics of such a quantum Markov chain
is confined to 
an
attractor spaces. This attractor space is spanned by all
orthogonal eigenspaces of the generating random unitary transformations
which are associated with eigenvalues of unit modulus.
Based on an analysis of these particular spectral properties
convenient representations for the asymptotic dynamics of this special class of
quantum Markov chains can be
derived. Secondly, it has been shown how an orthogonal basis of such an
asymptotic attractor space is determined by a set of linear equations involving
the Kraus operators specifying the generating random unitary quantum operation.
Taking advantage of the fundamental contraction property of this special class of unital
and trace preserving quantum operations recently parts of these investigations have
been generalized also to
quantum Markov chains which are generated either by trace preserving and unital \cite{Liu2011} or by
trace preserving and sub-unital quantum operations \cite{Golovkins2011}.
Although these generalizations demonstrate that
these quantum operations
can be diagonalized
on their asymptotic attractor spaces
they
still leave important questions open concerning, for example, the explicit construction of bases of the attractor spaces and of their
associated dual bases.

Despite these recent developments
it is still unclear to what extent similarly powerful results
apply to the most general and physically relevant
quantum Markov chains.
This is due to the fact that general quantum operations are not contracting so that the ideas underlying the proofs of
these previous results do not apply.
The main purpose of the paper is to close this gap and to
generalize these previous results to quantum Markov chains
which are generated by arbitrary quantum operations. We present such a generalization revealing that for establishing analogous results
the contraction property is not essential. The desired structural properties
can be established already from the fact that general quantum operations are trace non-increasing and completely positive. These properties
imply the validity of generalized Schwartz-inequalities \cite{Bhatia,Paulsen}.
Saturating these inequalities imposes important  structural constraints
on the asymptotic attractor space. It will be shown in the following
that a saturation of these generalized Schwartz-inequalities
is possible whenever
a strictly positive operator exists
which is contracted or left invariant by the generating quantum operation of a quantum Markov chain.
Our results also shed new light onto characteristic properties of
fixed points of quantum operations thereby generalizing
recent results on the theory of fixed points of quantum operations and of noiseless subsystems in quantum systems with finite-dimensional Hilbert spaces \cite{Arias2002,Kribs2003,Yuan2011}.

This paper is organized as follows.
In order to establish the basic notions
in Sec. \ref{SecII} known basic properties of completely positive quantum maps are summarized.
In Sec. \ref{SecIII}
general properties of the asymptotic dynamics of quantum Markov chains generated by
general quantum operations are characterized.
We show that for large numbers of iterations the dynamics is confined to an attractor space which is spanned by, in general, non-orthogonal (simple) eigenvectors of the generating quantum operation. Thus, on this attractor space also
these most general physically relevant quantum Markov chains are diagonalizable.
The determination of the asymptotic dynamics requires the projection of an arbitrary initially prepared
quantum state onto this attractor space. For this purpose one has to construct a
dual basis of this attractor space. In general, this is a complicated task in cases in which
the dimension of the underlying Hilbert space is large
even if
the dimension of the attractor space is small.
The construction of these dual bases
is addressed in Sec. \ref{SecIV}.
It is shown that whenever
a strictly positive linear operator exists which is contracted or left invariant by the generating quantum operation
it is possible to construct this dual basis in a straightforward
way from a knowledge of the basis of the attractor space.
In  Sec.\ref{SecV},
the relation between the basis of the attractor space and
Kraus operators defining the generating quantum operation of a quantum Markov chain is found.
Finally, in Sec.\ref{SecVI} it is demonstrated that to some extent
the theoretical treatment of Secs.\ref{SecIV} and \ref{SecV} can also be applied to cases in which the
quantum state which is contracted or left invariant by a quantum operation is not strictly positive.

\section{Basic properties of quantum operations\label{SecII}}
In this section known basic properties of completely positive quantum operations are summarized in order to introduce
our notation and in order to stress properties which are important
for our subsequent discussion.

In the following we consider a $N$-dimensional Hilbert space $\hil H$
equipped with a scalar product $(.,.)$. Let $\set B(\hil H)$ be the
associated Hilbert space of all linear operators acting on $\hil H$ with
the
Hilbert-Schmidt scalar product $\<A,B\>_{HS}=\Tr\{A^{\dagger}B\}$. The corresponding
Hilbert-Schmidt norm is given by $\parallel A \parallel := \sqrt{\<A,A\>_{HS}}$ for all $A \in \set B(\hil H)$.
Consequently, the induced norm of a
linear operator $\map S: \set B(\hil H)\rightarrow \set B(\hil H) $ acting on the space $\set B(\hil H)$ can be defined by
\begin{equation}
\label{def_induced_norm}
\parallel \map S \parallel = \sup_{\parallel A \parallel =1} \parallel \map S(A) \parallel.
\end{equation}
This latter norm fulfills the important relation (see e.g. \cite{Bhatia})
\begin{equation}
\label{eq_norm}
\parallel \map S \parallel = \parallel \map S^{\dagger} \parallel
\end{equation}
characterizing a Banach$^*$-algebra \cite{Bratteli}.
Thereby,
$\map S^{\dagger}$ denotes
the adjoint map of $\map S$ with respect to the Hilbert-Schmidt scalar product.
If a linear map satisfies the relation $\parallel \map S \parallel \leq 1$ it is called a contraction.

An arbitrary completely positive  linear
map $\map P: \set B(\hil H) \rightarrow \set B(\hil H)$ admits a
decomposition into Kraus operators $\{A_j\}_{i=1}^{k} \subseteq  \set B(\hil H)$ \cite{quantumoperations,Kraus-decomposition}, i.e.
\begin{equation}
\label{def:quantum_operation}
\map P\left(.\right) = \sum_{j=1}^{k} A_j \left(.\right) A_j^{\dagger}.
\end{equation}
Its adjoint map $\map P^{\dagger}$ is also a completely positive map with the
Kraus operators $\{A_j^{\dagger}\}_{i=1}^{k}$, i.e.
\begin{equation}
\label{eq:adjoin_QO}
\map P^{\dagger}\left(.\right) = \sum_{j=1}^{k} A_j^{\dagger} \left(.\right) A_j.
\end{equation}
In our subsequent discussion we call a
completely positive map $\map P$ which is also trace non-increasing, i.e.
$\map P^{\dagger}(I)=\sum_j A_j^{\dagger} A_j \leq I$, a quantum operation.
In the special case
$\map P^{\dagger}(I)=\sum_j A_j^{\dagger} A_j = I$ such a map is called a quantum channel or a
trace preserving quantum operation.
A quantum operation which leaves maximally mixed states undisturbed is called unital and satisfies the relation
$\map P(I)=\sum_j A_j A_j^{\dagger} = I$.
Prominent examples of unital quantum channels are random unitary operations (or random external fields) \cite{NovotnyJPA,NovotnyCEJP}.
In less restrictive cases in which $\map P(I)=\sum_j A_j A_j^{\dagger} \leq I$ quantum operations are called sub-unital.

In the subsequent sections the dynamics of a quantum system with Hilbert space $\hil H$ is discussed
which is governed by the iterated application of a quantum operation $\map P$ as described by
Eq.(\ref{def:quantum_operation}). This dynamics constitutes a quantum Markov chain with generator $\map P$ in analogy
to the corresponding classical case
(compare e.g. with Refs. \cite{quantumMarkovdefinition,Faigle2010}). Thus,
starting with an initial quantum state $0\leq \rho(0) \in \set B(\hil H)$,
after $n$ iterations this state is transformed into the quantum state
$\rho(n)=\map P^n(\rho(0))$.
Our main purpose is to analyze characteristic features of the resulting asymptotic behavior of $\map P^n$
and its relation to the spectral properties of its generating quantum operation $\map P$.
In general the commutation relation $\left[\map P,\map P^{\dagger}\right] = 0$ need not hold so that in general a diagonalization of
the generator
${\cal P}$ is not possible. Therefore, in such cases the determination of the resulting asymptotic dynamics of $\map P^n$
is complicated as this $n$-fold iteration may involve high powers of non-trivial Jordan normal forms of the generator $\map P$.

Despite these possible complications the following useful theorem holds which captures some basic spectral properties of the
special class of quantum channels, i.e. of trace preserving
completely positive quantum operations \cite{Bruzda2009,Faigle2010}.
\begin{theorem}
\label{theorem_basic_spectral_properties}
If $\map P$ is a trace preserving quantum operation of the form (\ref{def:quantum_operation}) and if
$\sigma$ denotes the set of all its eigenvalues the following statements are equivalent:
\begin{itemize}
\item[(1)] If $\lambda \in \sigma$ then $|\lambda| \leq 1$.
\item[(2)] $1 \in \sigma$.
\item[(3)] For every quantum state $0 \leq \rho \in \set B(\hil H)$ the limit
\begin{equation}
\overline{\rho} = \lim_{n \rightarrow \infty} \frac{1}{n} \sum_{k=0}^{n-1} \map P^{k}(\rho)
\end{equation}
exists and the quantum state $0\leq \overline{\rho}\in \set B(\hil H)$
is a fixed point and eigenvector of $\map P$ with eigenvalue $\lambda = 1$.
\end{itemize}
\end{theorem}

In our discussion
two concepts for analyzing characteristic properties of quantum operations ${\cal P}$
play an important role, namely
eigenspaces and their ranges.
Suppose
$\lambda$ is an eigenvalue of the map $\map P$. Its
corresponding eigenspace is denoted by
\begin{equation}
\set{Ker}(\map P- \lambda I)=\left\{X \in \mathcal{B}(\hil H)|
\map P(X)=\lambda X\right\} ,
\end{equation}
and the associated range of the map $\map P-\lambda I$ is denoted
by
\begin{equation}
\set{Ran}(\map P- \lambda I)=\left\{X \in \mathcal{B}(\hil H)|
\exists Y \in \mathcal{B}(\hil H), \hspace{0.5em}
X=\map P(Y)-\lambda Y\right\}.
\end{equation}

Recently, the asymptotic behavior of quantum Markov chains which are generated by
random unitary transformations has been investigated in detail
\cite{NovotnyJPA,NovotnyCEJP}. As a major result it has been shown that all Jordan blocks corresponding to eigenvalues
$\lambda$
with  $|\lambda|=1$ are one-dimensional and that the asymptotic dynamics of ${\cal P}^n$ can be diagonalized
on the associated attractor space
\begin{equation}
\set{Attr}(\map P) := \bigoplus_{|\lambda|=1}\set{Ker}(\map P- \lambda I)
\end{equation}
formed by the direct sum of all eigenspaces of ${\cal P}$ with eigenvalues $|\lambda|=1$.
The original idea of the proof of this asymptotic structure as presented in Ref.\cite{NovotnyCEJP}
relies on the fact that random unitary transformations
are contractions.
Recently, this result has been generalized to Markov chains generated by unital quantum channels \cite{Liu2011}
or by sub-unital quantum operations \cite{Golovkins2011}.
Although these investigations address  a large class of iterated quantum operations
the extension of these results to quantum Markov chains generated by arbitrary quantum operations, which are
generally non-contracting, is still unclear as the previously applied arguments do no longer apply.
In the following we discuss this issue and
generalize previous results to quantum Markov chains
generated by arbitrary quantum operations, i.e. completely positive and trace non-increasing quantum maps.

\section{Asymptotic dynamics of quantum Markov chains\label{SecIII}}
In this section the asymptotic dynamics of quantum Markov chains generated by arbitrary quantum operations is characterized.
In general, this characterization is complicated by the fact that in general the generating quantum operations are
not diagonalizable.

We show that for quantum operations, i.e. trace non-increasing
completely positive quantum maps
$\map P: \set B(\hil H) \rightarrow \set B(\hil H)$
(compare with
Eq.(\ref{def:quantum_operation})) with
$\map P^{\dagger}(I) \leq I$, in the limit $n \to \infty$ the dynamics
of the iterated map $\map P^n$ is confined to an attractor space
$\map Attr (\map P)$. This latter space is spanned by all, in general non-orthogonal
(simple), eigenvectors of $\map P$
with eigenvalues $\lambda$ of unit modulus, i.e. $|\lambda | =1$. Thus, on this asymptotically
relevant subspace $\map Attr(\map P)$
the iterated quantum operation $\map P^n$ can be diagonalized.
This main result is derived in two steps which are captured
by two main theorems.
The first theorem shows that the eigenvalues $\lambda$ of a quantum operation $\map P$
are confined to the unit circle, i.e. $|\lambda| \leq 1$, and that the eigenspaces of eigenvalues
from the set $\sigma_1 := \{\lambda \mid |\lambda| = 1\}$ are one dimensional but non-orthogonal, in general.
The second theorem exploits these properties and presents an explicit expression for the asymptotic dynamics
of an arbitrary initially prepared quantum state under the influence of the iterated quantum operation $\map P^n$.

\begin{theorem}
\label{theorem:III.1}
For any
quantum operation $\map P: \set B(\hil H) \rightarrow \set B(\hil H)$ as defined by
(\ref{def:quantum_operation}), i.e. for any trace non-increasing  and completely positive map,
the following relations hold:
\begin{itemize}
\item[(1)]  Any eigenvalue $\lambda$ of $\map P$ fulfills the relation $\mid \lambda\mid \leq 1$.
\item[(2)]  The kernel $\set{Ker}\left(\map P-\lambda I\right)$ and the range
$\set{Ran}\left(\map P-\lambda I\right)$ of any eigenvalue $\lambda$ with $\mid \lambda\mid~=~1$
fulfill the relation
\begin{equation}
\set{Ker}\left(\map P-\lambda I\right) \cap \set{Ran}\left(\map P-\lambda I\right) = \left\{0\right\}.
\end{equation}
\end{itemize}
\end{theorem}
\Proof{According to a general theorem \cite{Choi,Jam} a
linear map $\map P: \set B(\hil H) \rightarrow \set B(\hil H)$ is completely positive
iff the quantum state $\map I \otimes \map P(|\Phi\rangle \langle \Phi|)$ resulting from the
extended map $\map I \otimes \map P: \set B(\hil H)\otimes  \set B(\hil H) \rightarrow \set B(\hil H)\otimes \set B(\hil H)$
by acting on the entangled pure state
$|\Phi\rangle = 1/\sqrt{N}\sum_{i=1}^N|i\rangle |i\rangle \in \hil H \otimes \hil H$
is positive, i.e.
$\map I \otimes \map P(|\Phi\rangle \langle \Phi|)\geq 0$.
The positivity of this quantum state implies
\begin{eqnarray}
&&\mid \langle i|\langle l| \map I \otimes \map P(|\Phi\rangle \langle \Phi|)|j\rangle |k\rangle\mid ^2
= \mid \frac{1}{N} \langle l| \map P(|i\rangle \langle j|)|k\rangle \mid^2
\leq\nonumber\\
&&  \frac{1}{N} \langle l| \map P(|i\rangle \langle i|)|l\rangle
 \frac{1}{N} \langle k| \map P(|j\rangle \langle j|)|k\rangle
\label{ChoiJam}
\end{eqnarray}
with
$|i\rangle |l\rangle, |j\rangle |k\rangle$ denoting arbitrary elements of an orthonormal basis in the
extended Hilbert space $\hil H \otimes \hil H$.
As a consequence, the relations
\begin{eqnarray}
\sum_{l,k=1}^{N}\mid \langle l| \map P(|i \rangle \langle j|)|k\rangle \mid^2
&&\leq 1
\end{eqnarray}
are fulfilled for all possible values of $i,j$ if,
in addition, the
linear map $\map P$ is trace non-increasing.

Assuming that $\lambda$ is an eigenvalue of a completely positive and trace non-increasing quantum operation $\map P$ with its corresponding eigenvector $X$ we can conclude
\begin{eqnarray}
&&
\parallel \map P (X) \parallel^2 =
\mid \lambda \mid^2 \parallel X\parallel^2 =
\\
&&
\parallel
\sum_{i,j=1}^{N} \sum_{l,k=1}^N |k\rangle \langle l|
 \langle k|\map P(|i\rangle \langle j|)  |l\rangle \langle i|X|j\rangle
\parallel^2 =\nonumber\\
&&
\sum_{l,k=1}^N \mid \sum_{i,j=1}^N \langle i|X|j\rangle \langle k| \map P(|i\rangle \langle j|)  |l\rangle \mid^2
\leq
\nonumber\\ &&
 \sum_{i,j=1}^N \mid\langle i|X|j\rangle \mid^2
\underbrace{\sum_{l,k=1}^N
\mid \langle k| \map P(|i\rangle \langle j|)  |l\rangle \mid^2}_{\leq 1} \leq \parallel X \parallel^2\nonumber
\end{eqnarray}
with $\parallel A \parallel = \sqrt{\sum_{l,k=1}^N \mid \langle l|A|k\rangle \mid^2 }$
denoting
the Hilbert-Schmidt norm of a linear operator $A\in \set B(\hil H)$. Thus, we finally
arrive at the first part $(1)$ of this theorem, i.e. $|\lambda \mid \leq 1$.

For the second part {\it (2)} of this theorem let us
assume that this statement is false. Thus, there is an operator $0 \neq A \in \set B(\hil H)$ with
$A \in \set{Ker}\left(\map P-\lambda I\right) \cap \set{Ran}\left(\map P-\lambda I\right)$.
This implies $\map P(A)=\lambda A$ and there is an operator $0 \neq B \in \set B(\hil H)$
such that $\map P(B)=\lambda B + A$. By induction it can  be verified that
\begin{eqnarray}
\label{eq:induction}
\lambda^{n-1}A &=& \frac{1}{n}\map P^n(B) - \frac{1}{n} \lambda^n B
\end{eqnarray}
for all numbers of iterations $n\geq 1$.
As $n$ increases the second term on the right-hand side of (\ref{eq:induction})
becomes arbitrarily small in comparison with the term on the left hand side of (\ref{eq:induction})
for $|\lambda | = 1$.
Due to complete positivity of the map $\map P$
this is also true for the first term on the right hand side of (\ref{eq:induction}).
In order to demonstrate this
let us
consider an orthonormal basis $\{|i\>\}$ of the Hilbert space $\hil H$ so that the linear operators
$\{|i\>\<j|\}$ form an orthonormal basis of the space $\set B(\hil H)$. This implies the relation
\begin{equation}
\label{eq:P_decomposition}
\map P^n(B) = \sum_{i,j=1}^N B_{i,j} \map P^n(|i\>\<j|)
\end{equation}
with $B_{i,j}=\<i|B|j\>$. As the quantum operation $\map P^n$ is completely positive and
trace non-increasing
we can conclude from Eq.(\ref{ChoiJam}) that
\begin{equation}
\label{eq:elements_positivity}
0 \leq |\left[ \< l| P^n\left(|i\>\<j|\right)\right] |k\>| \leq 1
\end{equation}
for all $n\geq 0$ and for all $1 \leq k,l \leq N$.
So, in view of (\ref{eq:P_decomposition}) and (\ref{eq:elements_positivity}) also
all matrix elements of the first term
on the right-hand side of (\ref{eq:induction}) tend to zero for $|\lambda | = 1$
as $n$ tends to infinity. Thus,
we finally arrive at the contradiction $A=0$ to the initial assumption $A\neq 0$.$~\Box$.}

\begin{theorem}
\label{theorem:III.2}
Asymptotically
the iterative dynamics
of any quantum operation $\map P: \set B(\hil H) \rightarrow \set B(\hil H)$ is given by
\begin{eqnarray}
X_{\infty}(n) &=& \sum_{\lambda \in {\sigma}_{1},i=1}^{d_{\lambda}} \lambda^n X_{\lambda,i}
{\Tr}(X^{\lambda, i\dagger} X(0))
\label{asymptoticdynamics}
\end{eqnarray}
with
\begin{eqnarray}
\lim_{n\to \infty}
\parallel X_{\infty}(n) - \map P^n(X(0))\parallel &=& 0.
\end{eqnarray}
Thereby,
the eigenvectors $X_{\lambda,i}$
are determined by the relation $P(X_{\lambda,i}) = \lambda X_{\lambda,i}~~(i=1,...,d_{\lambda})$
with $\lambda \in \sigma_1 := \{\lambda | |\lambda | = 1\}$.
Their dual vectors $ X^{\lambda, i}\in \set B(\hil H)~~(i=1,...,d_{\lambda})$
with respect to the Hilbert-Schmidt scalar product
are defined by the property
$ {\Tr}(X^{\lambda, i\dagger}X_{\lambda',i'}) ~=~
\delta_{\lambda \lambda'}\delta_{ii'}$
for all $\lambda, \lambda' \in \sigma_1$ and by the property that for $\lambda \in \sigma_1$ each
$X^{\lambda,i}$ is orthogonal to all eigenspaces $\set{Ker}(\map P - \lambda' I)$ with $|\lambda'| <1$.
\end{theorem}
\Proof{Any linear map $\map P: \set B(\hil H) \rightarrow \set B(\hil H)$ can be brought into Jordan normal
form by an appropriate basis transformation $T$, i.e.
\begin{eqnarray}
 \map P &=& T J T^{-1}
\end{eqnarray}
with the (non singular) linear operator $T\in \set B(\hil H)$
defining the basis transformation
and with $J\in \set B(\hil H)$  denoting
the Jordan normal form of $\map P$.
Thus, $J$ is given by a direct sum of Jordan blocks $J_k$ of dimensions $d_k$, i.e.
$J = \sum_k \oplus J_k$. The matrix representation of any of these
$d_k\times d_k$-dimensional Jordan blocks
$J_k$ is given by
\begin{eqnarray}
 (J_k)_{ij} =
\left(
\begin{array}{ccccccc}
\lambda_k & 1& 0& 0& \cdots & 0&0\\
0&\lambda_k&1&0&\cdots &0&0\\
&&&\cdots&&\\
0&0&0&0&\cdots& 0&1\\
0&0&0&0&\cdots &0&\lambda_k
\end{array}
\right).
\end{eqnarray}
Consequently also the iterated map $\map P$ can be transformed to Jordan normal form, i.e.
$\map P^n = T^{-1}J^n T$. It is straightforward to demonstrate \cite{NAJ11} that the
modulus of the $(i,j)$-th matrix element of a Jordan block $(J_k)^n$ is given by
\begin{eqnarray}
 \mid (J_k)^n_{ij}\mid &=& \mid \lambda_k\mid ^{n-(j-i)}{n \choose n-(j-i)} \leq
\mid \lambda_k\mid^{n - d_{k}} n^{d_k}.\nonumber
\end{eqnarray}
Therefore, in the limit of large numbers of iterations $n$ the contributions of all Jordan blocks
with eigenvalues $\mid \lambda_k\mid < 1$ become vanishingly small so that in this limit only the Jordan blocks
with eigenvalues $\mid \lambda_k \mid = 1$ contribute to the iterated map $\map P^n$.

Let us now concentrate on these asymptotically contributing Jordan blocks $J_k$ with eigenvalues
$\mid \lambda_k \mid =1$. From theorem \ref{theorem:III.1} we can draw the conclusion that all these
Jordan blocks are one dimensional
so that the map $\map P^n$ can be diagonalized within the asymptotic subspace $\set{Attr}(\map P)$ which
is spanned by all eigenvectors $X_{\lambda, i},~i=1,...,d_{\lambda}$
with possibly $d_{\lambda}$-fold degenerate eigenvalues $\mid \lambda\mid =1$. These eigenvectors $X_{\lambda,i}$
fulfill the relations $\map P(X_{\lambda,i}) = \lambda X_{\lambda,i}$ for $i=1,...,d_{\lambda}$ and
$\lambda \in \sigma_1 := \{\lambda \mid |\lambda | = 1\}$ and in general they are not orthogonal (with respect
to the Hilbert-Schmidt scalar product). Consequently,
in the limit of large numbers of iterations $n$ it is sufficient to expand any initial linear operator
$X(0) \in \set B(\hil H)$ in terms of these eigenvectors, i.e.
\begin{eqnarray}
 X(0) &=& \sum_{\lambda \in \sigma_1} \sum_{i=1}^{d_\lambda} x_{\lambda,i} X_{\lambda, i} + Y(0)
\end{eqnarray}
with $Y(0)$ denoting the part of $X(0)$
which is contained in eigenspaces
of eigenvalues $\mid \lambda \mid < 1$. The coefficients $x_{\lambda, i}$ are given by
\begin{eqnarray}
 x_{\lambda, i} &=& {\Tr}(X^{\lambda,i\dagger}X(0))
\end{eqnarray}
with $X^{\lambda,i}\in \set B(\hil H)$ denoting the dual basis vector of $X_{\lambda, i}$ which fulfills
the relations ${\Tr}(X^{\lambda,i}X_{\lambda',i'}) = \delta_{\lambda \lambda'}\delta_{i i'}$ for all generalized eigenvectors
of a Jordan basis, i.e.
$|\lambda|, |\lambda'| \leq 1$ and $i=1,...,d_{\lambda}$, $i' = 1,...,d_{\lambda'}$.
As $\map P^n(Y(0))$ tends to zero in the limit $n\to \infty$ we finally obtain the
result of theorem \ref{theorem:III.2}.
$~\Box$

Theorem \ref{theorem:III.2} generalizes previous results on
the asymptotic dynamics of quantum Markov chains which apply only to
subunital channels and unital quantum operations or even to more restrictive cases,
such as unital channels, i.e. $\map P(I) =I$, $\map P^{\dagger}(I)\leq I$
or random unitary operations \cite{NovotnyJPA,NovotnyCEJP}.
It implies considerable simplifications
as far as the determination of the asymptotic dynamics is concerned
because only eigenspaces corresponding to eigenvalues
of unit modulus contribute and
these eigenspaces are associated with trivial one dimensional
Jordan blocks.
However, in general these
eigenspaces spanning the asymptotic attractor space $\set{Attr}(\map P)$ may still be
non-orthogonal. This complicates the
construction of the relevant dual vectors which project
onto the attractor space because for this purpose
typically the knowledge of the complete Jordan basis of generalized eigenvectors of the quantum operation $\map P$ is required.
This fact is summarized in the following corollary.
\begin{corollary}
Let $\{X_{\lambda,i}\}$ with $|\lambda|\leq 1$ and $i=1,...,d_{\lambda}$ be a complete Jordan basis of the quantum
operation $\map P$. The corresponding non-singular Hermitian matrix
\begin{eqnarray}
g_{\lambda i, \lambda'i'} &=& {\Tr}(X_{\lambda,i}^{\dagger} X_{\lambda',i'})
\end{eqnarray}
with $|\lambda|,|\lambda'|\leq 1$, $i=1,...,d_{\lambda}$, $i'=1,...,d_{\lambda'}$
contains all relevant information about the non-orthogonality of this Jordan basis.
The corresponding dual basis
$\{X^{\lambda,i}\}$ with $|\lambda|\leq 1$ and $i=1,...,d_{\lambda}$ is then given by
\begin{eqnarray}
X^{\lambda,i} &=& \sum_{|\lambda'|\leq 1, i'=1}^{d_{\lambda'}} (g^{-1})_{\lambda i,\lambda'i'} X_{\lambda' i'}.
\end{eqnarray}
In terms of this Jordan basis and its dual the projection operator $\Pi$ onto the attractor space $\set Attr(\map P)$ is given by
\begin{eqnarray}
 \Pi~.~ &=& \sum_{\lambda \in \sigma_1, i=1}^{d_{\lambda}} X_{\lambda, i}{\Tr}(X^{\lambda, i \dagger}~.~).
\end{eqnarray}
\end{corollary}

Thus, despite the simplifications resulting from theorem \ref{theorem:III.2},
in general the determination of the required dual vectors of the asymptotic attractor space $\set{Attr}(\map P)$
still constitutes a formidable task in particular in cases in which the dimension of the Hilbert space is large.
Nevertheless, in the subsequent sections it is
demonstrated that under additional restrictions on the generating quantum operations
$\map P$ of a quantum Markov chain both tasks, namely the construction of a basis for the asymptotic attractor space
$\set{Attr}(\map P)$ and
the construction of its associated dual basis, can be simplified considerably.

\section{Construction of the dual asymptotic basis\label{SecIV}}
A major open problem which has not been addressed in the previous section is whether there exist convenient methods which simplify
the
construction of the dual vectors
$ X^{\lambda, i}~~(i=1,...,d_{\lambda})$ for all possible eigenvalues $\lambda \in \sigma_1$. The knowledge of these
dual vectors is crucial for projecting any linear operator $X(0)$ or any initially prepared quantum state
$\rho (0)$ onto the attractor space $\set{Attr}(\map P)$ according to Eq.(\ref{asymptoticdynamics}).

In this section it is shown
that under the additional assumption, that the generating quantum operation $\map P$ of a quantum Markov chain  supports the existence of
a strictly positive quantum state $0 < \rho \in \set B(\hil H)$ which is contracted or left invariant, i.e.
$\map P(\rho) \leq \rho$, a straightforward construction of these dual basis vectors is possible from the knowledge of
all eigenvectors with eigenvalues $\lambda \in \sigma_1$.
For this construction we
exploit the basic property that the generating quantum operation $\map P$  of a quantum Markov chain is trace non-increasing and thus
fulfills characteristic
generalized Schwartz-inequalities
\cite{Bhatia,Paulsen}.
Saturating these inequalities we arrive at the following theorem.

\begin{theorem}
\label{theorem:IV.1}
If $\map P: \set B(\hil H) \rightarrow \set B(\hil H)$ is a
quantum operation
with the additional property that there exists a  quantum state $0 <  \rho  \in \set B(\hil H)$ fulfilling the
inequality
$\map P (\rho) \leq \rho$
then for all kernels of eigenvalues $\lambda \in \sigma_1$
the following equivalences hold:
\begin{itemize}
\item[(1)]  $X \in \set{Ker}(\map P - \lambda I)$ $\Leftrightarrow$ $X\rho^{-1} \in \set{Ker}(\map P^{\dagger} - (1/{\lambda}) I)$
\item[(2)]  $X \in \set{Ker}(\map P - \lambda I)$ $\Leftrightarrow$ $\rho^{-1}X \in \set{Ker}(\map P^{\dagger} - (1/{\lambda}) I)$
\item[(3)] $X \in \set{Ker}(\map P - \lambda I)$ $\Leftrightarrow$ $\rho^{-1}X\rho \in \set{Ker}(\map P - \lambda I)$.
\end{itemize}
Note that $\lambda \in \sigma_1$ implies the relation
$\overline{\lambda} = 1/\lambda$ with $\overline{\lambda}$ denoting the complex conjugate of $\lambda$.
\end{theorem}
\Proof{
For the proof of the first statement {\it (1)} we investigate the following linear map
\begin{equation}
\label{def_contraction}
\map V(X) = \map P^{\dagger}(X\rho^{-\frac{1}{2}})\rho^{\frac{1}{2}}
\end{equation}
with its adjoint map
\begin{equation}
\label{def_contraction}
\map V^{\dagger}(X) = \map P(X\rho^{\frac{1}{2}})\rho^{-\frac{1}{2}}.
\end{equation}
First of all we demonstrate
that both maps are contractions.
Because $\map P^{\dagger}$ is subunital, i.e. $\map P^{\dagger}(I) \leq I$, the
Schwartz operator inequality
$\map P^{\dagger}(X)\map P^{\dagger}(X^{\dagger}) \leq \map P^{\dagger}(XX^{\dagger})$ applies \cite{Bhatia,Paulsen}. Thus,
for a density operator $0 < \rho \in \set B(\hil H )$ which fulfills the relation
$\map P (\rho) \leq \rho$ we obtain the inequality
\begin{eqnarray}
\parallel \map V(X) \parallel^2 &=& \Tr\left\{(\map V(X))^{\dagger}\map V(X)\right\}
= \nonumber \\
&&\Tr\left\{\map P^{\dagger}(\rho^{-\frac{1}{2}}X^{\dagger})\map P^{\dagger}(X \rho^{-\frac{1}{2}})
\rho\right\}
\leq\nonumber\\
&&\Tr\left\{\map P^{\dagger}(\rho^{-\frac{1}{2}}X^{\dagger}X \rho^{-\frac{1}{2}})\rho \right\}
=\nonumber\\
&&
\Tr\left\{\rho^{-\frac{1}{2}}X^{\dagger}X \rho^{-\frac{1}{2}}\map P(\rho) \right\}
\leq\nonumber\\
&&\Tr\left\{\rho^{-\frac{1}{2}}X^{\dagger}X \rho^{-\frac{1}{2}}\rho \right\} =
\parallel X \parallel^2.
\end{eqnarray}
Consequently, $\parallel \map V \parallel = \parallel \map V^{\dagger} \parallel \leq 1$ and both
$\map V$ and $\map V^{\dagger}$ are contracting linear maps.

If $X \in \set{Ker}(\map P-\lambda I)$ a simple calculation reveals that
\begin{eqnarray}
\label{eq_1}
\parallel \map V^{\dagger}(X\rho^{-\frac{1}{2}}) \parallel &=&
\parallel \map P(X)\rho^{-\frac{1}{2}}\parallel =
\mid \lambda \mid \parallel X \rho^{-\frac{1}{2}} \parallel.
\end{eqnarray}
Furthermore,
Schwartz's inequality and the contracting properties of $\map V$ and $\map V^{\dagger}$ imply the inequalities
\begin{eqnarray}
&&
\parallel\map V^{\dagger}(X\rho^{-1/2})\parallel^2 = {\rm Tr}\left(
\left(\map V^{\dagger}(X\rho^{-1/2})\right)^{\dagger} \map V^{\dagger}(X\rho^{-1/2})
\right) = \nonumber\\
&&{\rm Tr}\left(
(X\rho^{-1/2})^{\dagger} \map V \map V^{\dagger}(X\rho^{-1/2})
\right) \leq\nonumber\\
&&
\parallel \map V \map V^{\dagger}(X\rho^{-1/2})\parallel
\parallel X\rho^{-1/2}\parallel\leq \nonumber\\
&&
\parallel \map V \parallel
\parallel
\map V^{\dagger}\parallel
\parallel (X\rho^{-1/2})\parallel^2 \leq
\parallel (X\rho^{-1/2})\parallel^2.
\end{eqnarray}
For $\lambda \in \sigma_1$ the very left hand side of these inequalities equals the very right hand side
so that we can conclude
\begin{equation}
\map V \map V^{\dagger}(X\rho^{-\frac{1}{2}})= \map P^{\dagger}(\map P(X)\rho^{-1}) \sqrt{\rho} = X\rho^{-\frac{1}{2}}
\end{equation}
for $\lambda \in \sigma_1$. Thus,
using $\map P(X)=\lambda X$ we finally arrive at the relation
\begin{equation}
\map P^{\dagger}(X\rho^{-1})= \frac{1}{\lambda}
X\rho^{-1} = \overline{\lambda}
X\rho^{-1}
\end{equation}
in view of $|\lambda |= 1$.
Analogously,
it can be demonstrated that
\begin{equation}
\map V^{\dagger} \map V (X\rho^{-\frac{1}{2}})=
\map P\left(
\map P^{\dagger}(X\rho^{-1}) \rho
\right)
\rho^{-1/2} =
X\rho^{-\frac{1}{2}}
\end{equation}
for $\lambda \in \sigma_1$,
so that we can conclude
$\map P(X) = \lambda X$ provided $\map P^{\dagger}(X\rho^{-1}) = \lambda^{-1} X\rho^{-1}$.

Statement {\it (2)} can be proven by applying the same reasoning to the linear map
$\map W(X)=\rho^{{1}/{2}}\map P^{\dagger}(\rho^{-{1}/{2}}X)$ and to its adjoint map
$\map W^{\dagger}(X)=\rho^{-1/2}\map P(\rho^{1/2}X)$.

Statement {\it (3)} is a simple consequence of statements {\it (1)} and {\it (2)}.
Assuming that $X$ is an eigenvector of $\map P$ with eigenvalue $\lambda \in \sigma_1$,
 i.e. $\map P(X)=\lambda X$,
statement {\it (2)} implies
$\map P^{\dagger}(\rho^{-1}X)=\frac{1}{\lambda}\rho^{-1}X$ and statement {\it (1)} implies
$\map P(\rho^{-1}X\rho)=\lambda \rho^{-1}X\rho$.$~\Box$
}

On the  basis of this theorem a new scalar product can be defined
in the space $\set B(\hil H)$. This scalar product is determined via any
strictly positive operator $0 < \rho\in \set B(\hil H)$ by  $\<A,B\>_{\rho} \equiv \<A,B\rho^{-1}\>_{HS}$.
It allows us to define the concept of $\rho$-orthogonality by the requirement
that two operators, say $A,B \in \set{B}(\hil H)$, are $\rho$-orthogonal, i.e. $A \bot_{\rho} B$,
iff $\<A,B\>_{\rho}=0$. Based on this concept
the following important $\rho$-orthogonality relations can be proved.

\begin{theorem}
\label{theorem:IV.2}
 Let $\map P: \set B(\hil H) \rightarrow \set B(\hil H)$ be a quantum operation
 and  let there be a strictly positive operator $0 <\rho \in  \set B(\hil H)$
such that $\map P(\rho) \leq \rho$,
then the following statements are fulfilled:
 \begin{itemize}
 \item[(1)] For any eigenvalue $\lambda$ of $\map P$ with $|\lambda|=1$ kernel and range are orthogonal, i.e.
 \begin{equation}
 \set{Ker}(\map P- \lambda I) \hspace{0.5em} \bot_{\rho} \hspace{0.5em} \set{Ran}(\map P- \lambda I)
 \end{equation}
and
\begin{equation}
\set{Ker}\left(\map P-\lambda I\right) \cap \set{Ran}\left(\map P-\lambda I\right) = \left\{0\right\}.
\end{equation}
 \item[(2)] For any two different eigenvalues
$\lambda_1$ and $\lambda_2$ of $\map P$ with $|\lambda_1|=|\lambda_2|=1$ the associated eigenspaces are orthogonal, i.e.
 \begin{equation}
 \set{Ker}(\map P- \lambda_1 I) \hspace{0.5em} \bot_{\rho} \hspace{0.5em} \set{Ker}(\map P- \lambda_2 I).
 \end{equation}
 \end{itemize}
\end{theorem}
\Proof{Let us consider $X \in \set B(\hil H)$ with $X \in \set{Ker}(\map P - \lambda I)$ and $\lambda \in \sigma_1$
 so that we obtain the relations
$\left( \map P(X)\right)^{\dagger} = \map P (X^{\dagger}) = \overline{\lambda} X^{\dagger}$ and
$\map P^{\dagger}(X \rho^{-1}) =
\frac{1}{{\lambda}} X \rho^{-1}$
from theorem \ref{theorem:IV.1}.
Furthermore, let us consider $Y\in \set B(\hil H)$ with $Y \in \set{Ran}(\map P - \lambda I)$, i.e.
there exists a $0 \neq Z \in \set B(\hil H)$ with $\map P(Z) - \lambda Z = Y$.
This implies the relation
\begin{eqnarray}
 &&\< X, Y \>_{\rho} = {\rm Tr}(X^{\dagger} Y \rho^{-1}) =
{\rm Tr}(X^{\dagger} \map P (Z) \rho^{-1}) - \\
&&
\lambda {\rm Tr}(X^{\dagger} Z\rho^{-1}) =
{\rm Tr}(\map [P^{\dagger}( X\rho^{-1})]^{\dagger}Z) -
\lambda {\rm Tr}(\rho^{-1} X^{\dagger} Z) = 0\nonumber
\end{eqnarray}
for $\lambda \in \sigma_1$.

In order to prove statement {\it (2)} let us consider
$X_1 \in \set{Ker}(\map P - \lambda_1 I)$ and
$X_2 \in \set{Ker}(\map P - \lambda_2 I)$ with
$\lambda_1 \neq \lambda_2$ and $\lambda_1,\lambda_2 \in \sigma_1$.
This implies the relation
\begin{eqnarray}
 &&\< X_1, X_2 \>_{\rho} = {\rm Tr}(X_1^{\dagger} X_2 \rho^{-1}) =
\frac{1}{\overline{\lambda}_1}
{\rm Tr}(\map P (X_1^{\dagger}) X_2 \rho^{-1}) =\nonumber\\
&&
\frac{1}{\overline{\lambda}_1}
{\rm Tr}(X_1^{\dagger} \map P^{\dagger}(X_2 \rho^{-1})) =
\frac{1}{\lambda_2\overline{\lambda}_1}
{\rm Tr}(X_1^{\dagger} X_2 \rho^{-1})) =\nonumber\\
&&
\frac{1}{\lambda_2\overline{\lambda}_1}
\<X_1,X_2 \>_{\rho}.
\end{eqnarray}
Because of $\lambda_1,\lambda_2 \in \sigma_1$ the relation
$\lambda_2 \overline{\lambda}_1 = \lambda_2/\lambda_1$ applies
so that $\<X_1,X_2\>_{\rho} = 0$ for $\lambda_1 \neq \lambda_2$.$~\Box$
}

Based on these characteristic properties the dual asymptotic basis can be constructed in a simple way
from the knowledge of all eigenspaces
$\set{Ker}(\map P - \lambda I)$ for all $\lambda \in \sigma_1$.
This central result of this section is summarized in the following theorem.
\begin{theorem}
\label{theorem:IV.3}
Let
$\map P: \set B(\hil H)\rightarrow \set B(\hil H) $
be a
quantum operation
with the additional
property that there exists a strictly positive quantum state $0 < \rho \in \set B(\hil H)$ with
$\map P(\rho) \leq \rho$. Under these conditions the dual vectors $X^{\lambda, i}$
of the eigenvectors
$X_{\lambda, i}$ with $\lambda \in \sigma_1$ and $i=1,...,d_{\lambda}$ which span the asymptotic
attractor space $\set{Attr}(\map P)$ are given by
\begin{eqnarray}
X^{\lambda,i} &=& X_{\lambda,i}~\rho^{-1} [{\Tr}(X_{\lambda,i}^{\dagger} X_{\lambda,i} \rho^{-1})]^{-1}.
\label{dualvectors}
\end{eqnarray}
\end{theorem}
\Proof{
The dimensions of the kernel and the range of the eigenspace of an arbitrary eigenvalue $\lambda$ fulfill
the general relation
\begin{eqnarray}
 Dim(\set{Ker}(\map P - \lambda I)) + Dim(\set{Ran}(\map P - \lambda I)) &\geq& Dim(\set B(\hil H)).\nonumber
\end{eqnarray}
According to theorem \ref{theorem:IV.2} the linear map $\map P$ fulfills the relation
$\set{Ker}(\map P - \lambda I)\perp_{\rho}\set{Ran}(\map P - \lambda I)$ so that we can conclude
\begin{eqnarray}
\set{Ker}(\map P - \lambda I)) \oplus \set{Ran}(\map P - \lambda I) &=&\set B(\hil H)
\end{eqnarray}
for all $\lambda \in \sigma_1$. In addition, theorem \ref{theorem:IV.2} also implies that
$\set{Ker}(\map P - \lambda_1 I)\perp_{\rho}\set{Ker}(\map P - \lambda_2 I)$ for all $\lambda_1 \neq
\lambda_2$ and $\lambda_1,\lambda_2 \in \sigma_1$ which implies the relation
\begin{eqnarray}
\bigoplus_{\lambda\in \sigma_1} \set{Ker}(\map P - \lambda I) \oplus
\bigcap_{\lambda \in \sigma_1}\set{Ran}(\map P - \lambda I) &=& B(\hil H)\nonumber
\end{eqnarray}
with $\bigcap_{\lambda \in \sigma_1}\set{Ran}(\map P - \lambda I)$ being orthogonal to the asymptotic
attractor space $\set{Attr}(\map P)$ and simultaneously containing all
contributions of eigenspaces
$\set{Ker}(\map P - \lambda I)$ with $|\lambda| <1$.
According to theorem \ref{theorem:III.2}
this latter subspace does not contribute to the dynamics of the iterated quantum operation $\map P^n$
in the limit $n \to \infty$. Therefore, $\set{Attr}(\map P)$ is spanned by the $\rho$-orthogonal
eigenspaces $\set{Ker}(\map P - \lambda I)$ with $\lambda \in \sigma_1$, i.e.
\begin{eqnarray}
 \set{Attr}(\map P) &=& \bigoplus_{\lambda\in \sigma_1} \set{Ker}(\map P - \lambda I)
\end{eqnarray}
and the $\rho$-orthogonal projection onto this attractor space is achieved by the dual vectors
of Eq.(\ref{dualvectors})
and by the projection operator
\begin{eqnarray}
\Pi\left(.\right) &=&
 \sum_{\lambda \in \sigma_1} X_{\lambda, i} {\rm Tr}(X^{\lambda,i\dagger}~.~).
\end{eqnarray}
Thereby, it is assumed that in case of degeneracy of eigenspaces with $\lambda \in \sigma_1$
the corresponding eigenstates $X_{\lambda,i}$ are $\rho$-orthogonalized.~~$\Box$

\section{Construction of the asymptotic attractor space\label{SecV}}
In this section we address the final question
how the relevant eigenvectors $X_{\lambda,i}~~i=1,...,d_{\lambda}$ with $\lambda \in \sigma_1$ which define the
asymptotic attractor space $\set{Attr}(\map P)$
are related to the Kraus operators which define the generating quantum operation of a quantum Markov chain.

For this purpose we use the
following property which holds for arbitrary completely positive maps} (\ref{def:quantum_operation})
even if they not trace non-increasing.
\begin{theorem}
\label{theorem:V.1}
Let $\map P: \set B(\hil H) \rightarrow \set B(\hil H)$ be a completely positive map defined by (\ref{def:quantum_operation})
and let there be a strictly positive operator $0 < \rho_1 \in \set B(\hil H)$ satisfying $\map P(\rho_1) \leq \rho_1$ and a
positive operator $\rho_2\geq 0$ satisfying $P^{\dagger}(\rho_2)\leq \rho_2$, then
any $X \in \set{Ker}(\map P - \lambda I)$ with $\lambda \in \sigma_1$
fulfills the set of equations
\begin{equation}
\rho_2 A_j X \rho_1^{-1} =\lambda \rho_2 X \rho_1^{-1} A_j
\end{equation}
for all $j \in \{1,\ldots,k\}$.
\end{theorem}
\Proof{In order to prove this theorem we investigate the linear map
\begin{equation}
V_j (X) =\lambda \sqrt{\rho_2}X\rho_1^{-1}A_j \sqrt{\rho_1} -
\sqrt{\rho_2}A_jX\rho_1^{-{1}/{2}}
\end{equation}
with $X \in \set{Ker}(\map P - \lambda I)$
and evaluate the following sum of non-negative terms
\begin{eqnarray}
\label{inequality_attractors}
0&\leq& \sum_{j=1}^{k}  \Tr\left\{V_j(X)[V_j(X)]^{\dagger}
\right\}
 = \Tr\left\{
X\rho_1^{-1}X^{\dagger}\map P^{\dagger}(\rho_2)
\right\} +
 \nonumber\\
&& |\lambda|^2 \Tr\left\{
X\rho_1^{-1}\map P(\rho_1)\rho_1^{-1}X^{\dagger}\rho_2
\right\}
- \nonumber \\
 && \lambda \Tr\left\{
X\rho_1^{-1}\map P(X^{\dagger})\rho_2
\right\}-
 \overline{\lambda} \Tr\left\{
\map P(X)\rho_1^{-1}X^{\dagger}\rho_2
\right\}
\leq  \nonumber \\
&&(1 - |\lambda|^2)\Tr\left\{
X \rho_1^{-1}X^{\dagger}\rho_2
\right\}.
\end{eqnarray}
Thereby,
the relations $\map P(X)=\lambda X$ and $\map P(X^{\dagger})=\overline{\lambda}X^{\dagger}$
together with the assumptions
$\map P(\rho_1) \leq \rho_1$ and
$P^{\dagger}(\rho_2)\leq \rho_2$
have been taken into account. For $\lambda \in \sigma_1$
this leads
to the conclusion $V_j(X)=0$ for
$X \in \set{Ker}(\map P - \lambda I)$ and for
each $j \in \{1,\ldots,k\}$.$~\Box$
}

Applying this theorem to a trace non-increasing quantum operation
$\map P$
and to its adjoint $\map P^{\dagger}$ (which is generally not trace non-increasing)
we obtain the following theorem.
\begin{theorem}
\label{theorem:V.2}
Let $\map P: \set B(\hil H) \rightarrow \set B(\hil H)$ be a quantum operation
and let there be a strictly
positive operator $0 < \rho \in \set B(\hil H)$ satisfying $\map P(\rho) \leq \rho$, then
any $X \in \set{Ker}(\map P - \lambda I)$ with $\lambda \in \sigma_1$
fulfills the set of equations
\begin{eqnarray}
A_j X \rho^{-1} &=& \lambda X \rho^{-1} A_j, \nonumber \\
A_j^{\dagger}X\rho^{-1} &=& (1/{\lambda})X\rho^{-1}A_j^{\dagger}, \nonumber \\
A_j \rho^{-1}X &=& \lambda \rho^{-1}X A_j, \nonumber \\
A_j^{\dagger}\rho^{-1}X &=& (1/{\lambda})\rho^{-1}X A_j^{\dagger}
\end{eqnarray}
for all $j \in \{1,\ldots,k\}$.
\end{theorem}
\Proof{
The first equation is a simple application of theorem \ref{theorem:V.1}
with $\rho_1=\rho$ and $\rho_2=I$. The second equation is a consequence of theorem
\ref{theorem:IV.1}, namely $\map P^{\dagger}(X\rho^{-1})=(1/{\lambda})X\rho^{-1}$,  and of
theorem \ref{theorem:V.1} applied to
$\map P^{\dagger}$ with $\rho_1=I$ and $\rho_2=\rho$.
The third and fourth equations are also consequences of these two theorems.
$~\Box$
}

This theorem states that for each eigenvalue $\lambda \in \sigma_1$ the set
\begin{eqnarray}
\set D_{\lambda,\rho} &\equiv&\left\{X\mid A_j X \rho^{-1} = \lambda X \rho^{-1} A_j, A_j^{\dagger}X\rho^{-1} = \overline{\lambda}X\rho^{-1}A_j^{\dagger},\right. \nonumber \\ && \qquad
A_j \rho^{-1}X = \lambda \rho^{-1}X A_j,
A_j^{\dagger}\rho^{-1}X = \overline{\lambda}\rho^{-1}X A_j^{\dagger} \nonumber \\ &&
\left. \qquad {\rm for}~j\in \{1,\ldots,k\} \right\}
\end{eqnarray}
includes the eigenspace $\set{Ker}(\map P - \lambda I)$, i.e.
$\set{Ker}(\map P-\lambda I) \subseteq \set{D}_{\lambda,\rho}$. In order to address the question under which conditions
we can achieve equality, i.e.
$\set{Ker}(\map P-\lambda I) = \set{D}_{\lambda,\rho}$, we add the following two corollaries of theorem \ref{theorem:V.2}
which state that either the conditions $\map P^{\dagger}(I) \leq I$ and $\map P(\rho) = \rho$
or the conditions $\map P^{\dagger}(I) = I$
and $\map P(\rho) = \rho$
are sufficient for this purpose.

\begin{corollary}
\label{corollary_unital_QO}
Let $\map P: \set B(\hil H) \rightarrow \set B(\hil H)$ be a 
quantum operation defined by (\ref{def:quantum_operation})
and let there be a strictly positive operator $0 < \rho \in \set B(\hil H)$ satisfying
$\map P(\rho) = \rho$, then the following relations hold:
\begin{itemize}
\item[(1)]
$\set{Ker}(\map P-\lambda I) = \set{D}_{\lambda,\rho}$
for all $\lambda \in \sigma_1$.
\item[(2)] If $X_1 \in \set{Ker}(\map P-\lambda_1 I)$, $X_2 \in \set{Ker}(\map P-\lambda_2 I)$ then $X_1X_2\rho^{-1} \in \set{Ker}(\map P-\lambda_1\lambda_2 I)$.
\end{itemize}
\end{corollary}
\Proof{For the proof of {\it (1)}
we have to show that $\set{D}_{\lambda,\rho} \subseteq \set{Ker}(\map P-\lambda I)$.
For this purpose let us assume that $X \in \set{D}_{\lambda,\rho}$ and therefore
satisfies the relations $A_j X \rho^{-1} = \lambda X \rho^{-1} A_j$ for $j\in \{1,...,k\}$.
Multiplying both sides of these equations by $\rho A_j^{\dagger}$ and summing over all values of $j=1,...,k$
we obtain the result $\map P(X)=\lambda X\rho^{-1}\map P(\rho)=\lambda X$, i.e. $X \in \set{Ker}(\map P-\lambda I)$.

If $X_1 \in \set{Ker}(\map P-\lambda_1 I)$ and  $X_2 \in \set{Ker}(\map P-\lambda_2 I)$
statement {\it (1)}  ensures  that $X_1\in \set{D}_{\lambda_1,\rho}$ and $X_2\in \set{D}_{\lambda_2,\rho}$. Thus,
from theorem \ref{theorem:IV.1} we can conclude that
$\rho X_2 \rho^{-1} \in \set{D}_{\lambda_2,\rho}$. Consequently, the following two identities are fulfilled
\begin{eqnarray}
A_jX_1X_2(\rho^{-1})^2 &=& A_jX_1\rho^{-1}\rho X_2(\rho^{-1})^2 \nonumber
\\ &=& \lambda_1X_1\rho^{-1}A_j\rho X_2 \rho^{-1} \rho^{-1} \nonumber \\ &=& \lambda_1\lambda_2 X_1 X_2 (\rho^{-1})^2A_j,
\nonumber\\
A_j \rho^{-1}X_1 X_2 \rho^{-1} &=& \lambda_1\rho^{-1}X_1 A_j X_2 \rho^{-1} \nonumber \\ &=& \lambda_1\lambda_2 \rho^{-1}X_1 X_2 \rho^{-1}
\end{eqnarray}
and analogous equalities for the adjoint linear operators $A_j^{\dagger}$ so that we can conclude
that $X_1 X_2 \rho^{-1} \in \set{D}_{\lambda_1\lambda_2,\rho}=\set{Ker}(\map P-\lambda_1\lambda_2 I)$.
$\Box$
}

Unital quantum operations are examples of quantum operations fulfilling the assumptions of this theorem.

This corollary generalizes previous
work on the theory of fixed points and noiseless subsystems of unital channels in Hilbert spaces of finite dimensions
\cite{Yuan2011,Arias2002,Kribs2003}. Firstly, these results generalize previously developed procedures
for evaluating asymptotically relevant eigenspaces of quantum Markov chains, which apply to the special eigenvalue
$\lambda =1$ only, to all asymptotically relevant eigenvalues $\lambda \in \sigma_1$. Secondly, this result applies not only to unital channels. Indeed, according to \ref{corollary_unital_QO} only one restriction is required for obtaining not only a necessary but also
a sufficient condition for the construction of $\set{Ker}(\lambda - I)$ for all $\lambda \in \sigma_1$, namely
$\map P(\rho)=\rho$ for some strictly positive $0< \rho \in \set B(\hil H)$.

\section{Generalizations to not strictly positive quantum states}\label{SecVI}
In our previous considerations we required the existence of a strictly positive quantum state such that
$\map P(\rho) \leq \rho$. However, there are quantum operations which do not fulfill this condition.
Therefore, the natural question arises whether some of our previous results also apply to such situations.

According to theorem \ref{theorem_basic_spectral_properties} a quantum channel $\map P$, i.e. a trace preserving
completely positive map, is always
equipped with a state $\rho$ such that $\map P(\rho) = \rho$. Although this state
satisfies the condition $\map P(\rho) \leq \rho$ it
need not be strictly positive. Let
\begin{equation}
\rho = \sum_{i} \alpha_i |\psi_i\>\<\psi_i|
\end{equation}
be its diagonal form with the orthonormal pure states $|\psi_i\>$ and $\alpha_i > 0$.
The orthogonal projection $P_{\rho} = \sum_{i}|\psi_i\>\<\psi_i|$
onto the support (or range) of this state $\rho$ satisfies the inequality
\begin{equation}
\map P(P_{\rho}) \leq \frac{1}{\alpha_{min}}\map P(\rho) \leq \frac{1}{\alpha_{min}\alpha_{max}}\map P(P_{\rho})
\end{equation}
with $\alpha_{min}$ (resp. $\alpha_{max}$) denoting
the minimal (resp. maximal) nonzero eigenvalue of the state $\rho$.
According to Ref.\cite{Krohn1978} the projection $P_{\rho}$ always reduces a quantum operation $\map P$,
i.e. $\map P(P_{\rho}XP_{\rho})=P_{\rho}\map P(P_{\rho}XP_{\rho})P_{\rho}$.
Hence the corresponding reduced map $\map P: \set B(P_{\rho}\hil H) \rightarrow \set B(P_{\rho}\hil H)$
constitutes a well defined quantum operation.
Indeed, complete positivity is apparent and if the original map $\map P$ is trace non-increasing (resp. trace preserving)
then also its reduction is trace non-increasing (resp. trace preserving).
Moreover, the fixed state $\rho$ is strictly positive on $P_{\rho}\hil H$. Hence, all requirements
for the applicability of our theorems from sections
\ref{SecIII} and \ref{SecIV}
are met for the restriction of a quantum channel $\map P$ onto the subalgebra $\set B(P_{\rho}\hil H)$.

This construction is feasible for an arbitrary not necessarily strictly positive
state $\rho$ satisfying $\map P(\rho) \leq \rho$.
Furthermore, there is always a maximal state $\tilde{\rho}$ satisfying this property.
"Maximal" means that $P_{\sigma} \leq P_{\tilde{\rho}}$ for an arbitrary state $\sigma$ satisfying
$\map P(\sigma) \leq \sigma$. In this sense the
orthonormal projection $P_{\tilde{\rho}}$ on the support of $\tilde{\rho}$ defines the so-called
recurrent subspace $P_{\tilde{\rho}}\hil H$ \cite{Krohn1978}, the maximal Hilbert space for which our theory applies.

\section{Summary and Conclusions}
We have investigated the
asymptotic dynamics of quantum Markov chains generated by
general quantum operations, i.e. by completely positive and trace non-increasing linear maps acting in a Hilbert space.
It has been shown that their resulting asymptotic dynamics is confined to
attractor spaces spanned by typically non-orthogonal eigenspaces  of the generating quantum operations associated
with eigenvalues of unit modulus.
These eigenvalues have trivial Jordan blocks so that asymptotically also these most general and physically relevant
quantum Markov chains are
diagonalizable on their
attractor spaces. Furthermore,
provided a strictly positive operator can be found, which is contracted or left invariant by the generating quantum operation
of such a quantum Markov chain,
an explicit construction of a basis of the attractor space and of its associated dual basis has been presented.
The basis vectors of the attractor space are determined by linear equations which depend
in a simple way on the Kraus operators defining the generating quantum operation. Each of these basis vectors
is related to its dual one by a linear transformation. This linear transformation is
defined by the inverse of the positive operator which is contracted or left invariant by the generating quantum operation.
This explicit construction of the asymptotic dynamics of an arbitrary iterated quantum operation is expected to
offer significant advantages whenever the dimension of the Hilbert space is large and at the
same time the dimension of the attractor space is small. Thus, the theoretical description of the
asymptotic dynamics described here may be particularly useful for applications in large quantum systems.

Apart from possible practical advantages our discussion also explicitly demonstrates the close connection between the
existence of a quantum state which is contracted or even left invariant
by the generating quantum operation of a quantum Markov chain
and the resulting asymptotic dynamics. This way it generalizes
recent results on the theory of fixed points of quantum operations and of noiseless subsystems
in quantum systems with finite-dimensional Hilbert spaces \cite{Arias2002,Kribs2003,Yuan2011}.

Finally, we want to emphasize again that
the results presented here are restricted to finite dimensional Hilbert spaces.
This is mainly due to arguments involved in the proof of theorem \ref{theorem:III.1}.
They involve a general theorem due to Choi \cite{Choi} and Jamio\l kowski \cite{Jam}
which applies to finite dimensional Hilbert spaces only and which relates
the complete positivity of a linear map to the positivity of an extended map.
However, as most
arguments involved in the proofs of our subsequent theorems also apply to infinite dimensional Hilbert spaces it is expected
that at least parts of the construction methods for attractor spaces presented here can be extended to infinite dimensional
Hilbert spaces.

\end{document}